\documentclass[preprint,12pt]{elsarticle}

\usepackage{amssymb}

\usepackage{color}

\journal{Applied Mathematical Modelling}

\begin{document}

\begin{frontmatter}



\title{Formation of sand ripples under a turbulent liquid flow \tnoteref{label_note_copyright} \tnoteref{label_note_doi}}

\tnotetext[label_note_copyright]{\copyright 2016. This manuscript version is made available under the CC-BY-NC-ND 4.0 license http://creativecommons.org/licenses/by-nc-nd/4.0/}

\tnotetext[label_note_doi]{Accepted Manuscript for Applied Mathematical Modelling, v. 39, p. 7390-7400, 2015, http://dx.doi.org/10.1016/j.apm.2015.03.021}


\author{Erick de Moraes Franklin}

\address{Faculdade de Engenharia Mec\^anica - Universidade Estadual de Campinas\\
e-mail: franklin@fem.unicamp.br\\
Rua Mendeleyev, 200 - Campinas - SP - CEP: 13083-970\\
Brazil}

\begin{abstract}
Sand ripples are commonly observed in both nature and industry. For example, they are found on riverbeds and in oil pipelines that transport sand. In both natural and industrial cases, ripples increase friction between the bed and fluid and are related to flooding, high pressure drops, and transients. Ripples appear when sediments are entrained as bed load (a mobile granular layer) and are usually considered to be the result of initial bedforms that eventually saturate. Given the small aspect ratio of the initial bedforms, linear analyses can be used to understand the formation of ripples. This paper presents a linear stability analysis of a granular bed under a turbulent flow of a liquid. This analysis takes into consideration all the main mechanisms and parameters involved in the turbulent liquid case, including some important parameters that have not yet been considered together such as the bed compactness and the bed-load threshold shear stress. The results of this analysis are compared with published experimental results and they show good agreement.
\end{abstract}

\begin{keyword}
Sediment transport \sep bed load \sep turbulent flow \sep instability \sep ripples

\end{keyword}

\end{frontmatter}



\section{Introduction}

Turbulent flows of liquids in the presence of sand are frequent in both nature and industry. Some examples of such flows include  river flows, ocean flows, and oil-water-sand flows in petroleum pipelines. Under moderate shear stresses, a granular bed is formed that is not fluidized by the liquid, and the sand is carried as a mobile layer in contact with the fixed part of the sand bed. This mode of transport is known as bed load.

A granular bed entrained as bed load may induce the formation of ripples and dunes \cite{Bagnold_1}. Ripples are bedforms whose wavelengths scale with the grain diameter but not with the flow depth \cite{Raudkivi_1, Engelund_Fredsoe}. They are usually considered to be a result of initially two-dimensional bedforms that saturate eventually \cite{Franklin_4, Franklin_5}. Dunes are bedforms whose wavelengths scale with both the flow field and the flow depth \cite{Raudkivi_1, Engelund_Fredsoe} and may be considered as the result of coalescence of ripples \cite{Raudkivi_Witte, Raudkivi_2, Fourriere_1, Franklin_6}.  Ripples and dunes increase friction between the bed and fluid and are related to flooding, high pressure drops, and transients. Although of importance, the formation of aquatic ripples has not yet been completely understood.

Many studies have been conducted in the past decades on the stability of granular beds sheared by fluids, most of them employing linear stability techniques \cite{Kennedy, Reynolds, Engelund_1, Fredsoe_1, Engelund_Fredsoe, Richards, Elbelrhiti, Charru_3, Claudin_Andreotti}. Although this approach has been criticized in the case of aquatic dunes \cite{Fourriere_1, Franklin_6}, it is justified in the case of aquatic ripples given the small aspect ratio of the initial bedforms from which ripples are formed.

More recently, Franklin \cite{Franklin_4} described the main mechanisms of ripple formation and presented a linear stability analysis for the specific case of turbulent liquid flows far from the threshold for grain displacement \cite{Bagnold_1, Raudkivi_1}. The analysis gave expressions for the wavelength, growth rate, and celerity of initial bedforms and demonstrated their variation with fluid stresses, grain diameters, and local slope. However, the analysis neglected the effects of the bed-load threshold shear stress, bed compactness, and settling velocity on bed stability. To the best of the author's knowledge, a complete linear analysis that considers all the physical effects (including the threshold effects and bed compactness) has not yet been performed.

Franklin \cite{Franklin_5} presented a nonlinear stability analysis within the same scope as one of his previous papers \cite{Franklin_4}. This nonlinear analysis, based on the weakly nonlinear approach \cite{Landau_Lifshitz}, showed that the initial bed instabilities saturate, because of which the wavelengths of ripples can be predicted by linear analyses even if ripples observed in nature are in a nonlinear regime.

This paper addresses the formation of sand ripples under turbulent flows of liquids, and presents a linear stability analysis that predicts the growth rate, celerity and length scale of the initial instabilities from which ripples are formed. The present analysis is different from previous works in that it considers all the main physical effects, including the threshold shear stress for grain displacement and bed compactness.
 
Section \ref{section:linear_stability} describes the physics and main equations involved in the linear stability analysis. Section \ref{section:results} presents the results of the stability analysis and compares it with published experimental results. Section \ref{section:conclusions} concludes the paper.

\section{Linear Stability}
\label{section:linear_stability}

Franklin \cite{Franklin_4} presented a linear stability analysis of a granular bed sheared by a turbulent liquid flow, without free-surface effects. The absence of a free surface is justified in the case of ripples, as these forms do not scale with the flow depth. Franklin's analysis was based on four equations, which describe the mass conservation of granular matter, fluid flow perturbation caused by the bed shape, the transport of granular matter by a fluid flow, and the relaxation effects related to the transport of grains. Although Franklin \cite{Franklin_4} presented the main mechanisms of ripple formation, he neglected the effects of the threshold shear stress for grain displacement, the bed compactness, and the settling velocity of grains on bed stability.

The linear model presented next is derived from that presented in Franklin \cite{Franklin_4}. This model is constructed in two dimensions, which is justified by Squire's theorem \cite{Drazin_Reid}. The model considers all the main mechanisms and effects involved in ripple formation; therefore, the stability analysis is more comprehensive than previous ones. The main equations, including those of Franklin \cite{Franklin_4}, are presented next for completeness of the paper; their development is reported in detail in Franklin \cite{Franklin_4}.

\subsection{Conservation equations}

The conservation equations used in this analysis are the mass conservation of granular matter and the momentum balance between the liquid and the grains. The mass conservation of grains in two dimensions relates the local height of the bed, $h$, to the local transport rate (volumetric) of grains per unit width $q$:

\begin{equation}
\frac{\partial h}{\partial t}\,+\,\frac{1}{\phi}\frac{\partial q}{\partial x}\,=\,0,
\label{eq:exner}
\end{equation}

\noindent where $t$ is the time, $x$ is the longitudinal direction and $\phi$ is the bed compactness. The momentum balance between the liquid and the grains is usually obtained by dimensional analysis, as there is no consensus about the rheology of granular matter. Semi-empirical momentum balances between the liquid and the grains were proposed in the previous decades, and the obtained expressions relate the bed-load transport rate to the shear stress caused by fluid flow on a granular bed. The expression proposed by Meyer-Peter and M\"{u}ller \cite{Mueller}, one of the most frequently used transport rate equations, is based on data from exhaustive experiments, and for this reason, it is used in the present model. The volumetric transport rate of grains per unit width $q_{0}$ for a fully developed flow \cite{Mueller} is given by

\begin{equation}
q_{0}\,=\,D_1 \left( \tau_0-\tau_{th}\right)^{3/2},
\label{eq:qsat_0}
\end{equation}

\noindent where $\tau_0$ is the shear stress on the granular bed caused by the fully developed flow (the unperturbed, basic state flow described next) and $\tau_{th}$ is the threshold shear stress for the incipient motion of grains \cite{Bagnold_1}. $D_1$ is given by:

\begin{equation}
D_1\,=\,\frac{8}{\rho^{3/2}\left[ \left( S-1\right) g\right] },
\label{eq:D1}
\end{equation}

\noindent where $\rho$ is the specific mass of the liquid, $S=\rho_p/\rho$, $\rho_p$ is the specific mass of the grain material and $g$ is the acceleration of gravity. According to Eq. \ref{eq:D1}, $D_1$ is constant for given fluid and grain types.

There is not a universal expression for the bed-load transport rate; on the contrary, there are several empirical and semi-empirical laws, many of them with the same functional form of the Meyer-Peter and M\"{u}ller \cite{Mueller} equation. Usually, the only difference is a multiplicative prefactor. Because the transport rate equations are semi-empirical laws, experimental uncertainties are included in the multiplicative prefactor. In addition, high uncertainties are present in the determination of the threshold shear stress $\tau_{th}$ because the threshold for incipient motion depends on the surface density of the moving grains. Charru et al. (2004) \cite{Charru_1} showed that the surface density of moving grains decays due to an increase in bed compactness, caused by the rearrangement of grains, known as armoring, which leads to an increase in the threshold shear rate for the bed load. The adoption of given values for the multiplicative prefactor and for the threshold Shields number, $\theta_{th}\,=\,\tau_{th} \left[\left( \rho_{p} - \rho \right) gd\right]^{-1}$, does not change the conclusions of the following analysis.

\subsection{Basic state}

The basic state corresponds to a fully-developed turbulent boundary layer over a flat granular bed. For a two-dimensional boundary layer in a hydraulic rough regime, the fluid velocity profile is given by \cite{Schlichting_1}

\begin{equation}
u\,=\,\frac{u_*}{\kappa}\ln\left(\frac{y}{y_0}\right),
\label{eq:u_basic}
\end{equation} 

\noindent where $u(y)$ is the longitudinal component of the mean velocity, $\kappa\,=\,0.41$ is the von K\'arm\'an constant, $y$ is the vertical distance from the bed, $y_0$ is the roughness length and $u_*=\rho^{-1/2}\tau_0^{1/2}$ is the shear velocity. 

In the steady state regime without spatial variations (flat bed), the fully developed fluid flow is given by Eq. \ref{eq:u_basic}. In this case, the fluid flow and the flow rate of grains are in equilibrium. This means that the available momentum for transporting grains as bed load is limited because part of the fluid momentum is transferred to the moving grains, which in turn transfer a part of the transferred fluid momentum to the fixed layers of the granular bed, until an equilibrium condition is reached. The equilibrium transport rate of grains is called \textit{saturated transport rate} and is given by Eq. \ref{eq:qsat_0}.

\subsection{Perturbations}

The origin of perturbations in this problem is the initial undulation of the granular bed. The bed undulation causes deviations from the basic state in both the fluid flow and the bed-load transport rate, thereby perturbing them. If the bed undulation is of a small aspect ratio, as is expected for initial instabilities, the perturbations in the fluid flow and in the transport rate may be assumed as being small and a linear model can be used. The equations for these perturbations are presented next.

In the previous decades, many analytical works were focused on the perturbation of a turbulent boundary layer by bedforms. Among them, we cite Jackson and Hunt \cite{Jackson_Hunt},  Hunt et al. \cite{Hunt_1}, and Weng et al. \cite{Weng} here. Sauermann \cite{Sauermann_2} and Kroy et al. \cite{Kroy_A,Kroy_C} simplified the results of Weng et al. \cite{Weng} for surface stress and obtained an expression containing only the dominant physical effects of the perturbation. For a hill with local height $h$ and a length $2L$ between the half-heights, they obtained the perturbation of the longitudinal shear stress (dimensionless):

\begin{equation}
\hat{\tau}_{k}=Ah(|k|+iBk),
\label{eq:stress_pert_fourier}
\end{equation}

\noindent where $A$ and $B$ are considered as constants, $k=2\pi\lambda^{-1}$ is the longitudinal wavenumber ($\lambda$ is the wavelength), and $i=\sqrt{-1}$. The shear stress on the bed surface is given as

\begin{equation}
\tau\,=\,\tau_{0}(1\,+\,\hat{\tau}).
\label{eq:stress_total}
\end{equation}

When the fluid flow is perturbed by the undulated bed, the bed-load transport rate varies locally. If equilibrium is assumed between the fluid flow and the bed-load transport rate, which is equivalent to neglecting the inertia of grains, then the transport rate is locally saturated and obtained by replacing $\tau_0$ in Eq. \ref{eq:qsat_0} with $\tau$ (Eq. \ref{eq:stress_total}). A convenient way to express this perturbed-saturated transport rate $q_{sat}$ is:

\begin{equation}
\frac{q_{sat}}{q_0}\,=\,\left( 1+D_2\tau\right)^{3/2},
\label{eq:qsat}
\end{equation}

\noindent where $D_2\,=\,\tau_0\left( \tau_0-\tau_{th}\right)^{-1}$ is a term that quantifies how far the flow is from the threshold. This term has a singularity at $\tau_0 = \tau_{th}$. Fourri\`ere et al. (2010) \cite{Fourriere_1} proposed a more sophisticated expression for the perturbed-saturated transport rate, however, the form used here (Eq. \ref{eq:qsat}) is simpler while allowing to analyze threshold effects.

In the case of a spatially varying perturbed flow, a relaxation effect exists between the fluid and the grains owing to the inertia of the latter \cite{Andreotti_2}. For this reason, the bed-load transport rate will lag behind the fluid flow by a certain distance,  which is usually referred to as \textit{saturation length}, $L_{sat}$. Andreotti et al. \cite{Andreotti_2} proposed the following expression taking into account the relaxation effect:

\begin{equation}
\frac{\partial q}{\partial x}\,=\,\frac{q_{sat}-q}{L_{sat}}.
\label{eq:relax}
\end{equation} 

For the specific case of bed load under liquid flows, Charru (2006) \cite{Charru_3} proposed that the saturation length $L_{sat}$ is proportional to a deposition length $l_d=\frac{u_*}{U_s}d$:

\begin{equation}
L_{sat}\,=\,C_{sat}\frac{u_*}{U_s}d,
\label{eq:lsat}
\end{equation}

\noindent where $U_s$ is the settling velocity of a single grain and $C_{sat}$ is a constant of proportionality. The saturation length given by Eq. \ref{eq:lsat} is experimentally supported by Franklin and Charru (2011) \cite{Franklin_8}.

P\"ahtz et al. (2013) \cite{Pahtz_1} and P\"ahtz et al. (2014) \cite{Pahtz_2} proposed a more general saturation length, suitable for both liquid and gas flows. The proposed expression is also supported by experimental measurements, including the results of Franklin and Charru (2011) \cite{Franklin_8}. However, as the present paper concerns only liquid flows, Eq. \ref{eq:lsat} is preferred.

Another parameter affecting bed stability is the local slope of the bed: the gravitational field weakens the transport of grains over positive slopes. One simple way to take into account this effect is to compute the effective shear stress perturbation by replacing $B$ in Eq. \ref{eq:stress_pert_fourier} with $B_e=B-B_g/A$ so that the perturbed stress takes into consideration the grain weight and the shear between the grains, in addition to the shear caused by the fluid flow \cite{Charru_3}.

\subsection{Solution}

Taking into account that the initial instabilities are of a small aspect ratio, solutions $h$ and $q$ to Eqs. \ref{eq:exner}, \ref{eq:stress_total}, \ref{eq:qsat}, and \ref{eq:relax} are plane waves. They can be decomposed into their normal modes as follows:

\begin{equation}
h(x,t)\,=\,H e^{i\left( kx -\Omega t\right)} + c.c.,
\label{eq:normal_h}
\end{equation}

\begin{equation}
\frac{q(x,t)}{q_0}\,=1+Q e^{i\left(kx -\Omega t\right)} + c.c.,
\label{eq:normal_q}
\end{equation}

\noindent where $k\in \mathbb{R}$, $k=2\pi\lambda^{-1}$ is the wavenumber in the $x$ direction, $\lambda \in \mathbb{R}$ is the wavelength in the $x$ direction, $H\in \mathbb{C}$ and $Q\in \mathbb{C}$ are the amplitudes, and $c.c.$ denotes ``complex conjugate''. Let $\Omega\in \mathbb{C}$, $\Omega\,=\,\omega\,+\,i\sigma$, where $\omega\in \mathbb{R}$ is the angular frequency and $\sigma\in \mathbb{R}$ is the growth rate. Inserting the normal modes in Eqs. \ref{eq:exner}, \ref{eq:stress_total}, \ref{eq:qsat}, and \ref{eq:relax} gives the following system:

\begin{equation}
\left[\begin{array}{cc}\sigma-i\omega & (1/\phi)ikq_0 \\ (3/2)D_2A\left( |k|+iB_ek \right) & -(1+ikL_{sat}) \\ \end{array}\right]\left[\begin{array}{c}H \\ Q \\ \end{array}\right] \,=\,\left[\begin{array}{c}0 \\ 0 \\ \end{array}\right].
\label{eq:eigenvalue}
\end{equation}

The non-trivial solution of Eq. \ref{eq:eigenvalue} gives the growth rate (Eq. \ref{eq:sigma}) and the angular frequency of initial instabilities. The phase velocity is given by $c\,=\,\omega /k$ (Eq. \ref{eq:c}).

\begin{equation}
\sigma\,=\,\frac{3}{2}\frac{Aq_0D_2}{\phi}\frac{k^2\left( B_e-|k|L_{sat}\right)}{1+\left( kL_{sat}\right)^2},
\label{eq:sigma}
\end{equation}

\begin{equation}
c\,=\,\frac{3}{2}\frac{Aq_0D_2}{\phi}\frac{|k|\left( 1+ B_eL_{sat}|k| \right)}{\left( 1+\left( kL_{sat}\right)^2\right)^2}.
\label{eq:c}
\end{equation}

The most unstable mode corresponds to the maximum of the growth rate. Considering $\partial\sigma /\partial k=0$ and within a long-wave approximation, i.e., $\left( kL_{sat}\right)^2 \ll 1$, the most unstable wavenumber $k_{max}$, wavelength $\lambda_{max}$, growth rate $\sigma_{max}$, and celerity $c_{max}$ are, respectively, given as

\begin{equation}
k_{max}\,\approx\,\frac{2}{3}\frac{B_e}{L_{sat}},
\label{k_max}
\end{equation}

\begin{equation}
\lambda_{max}\,\approx\,\frac{3\pi }{B_e}L_{sat},
\label{L_max}
\end{equation}

\begin{equation}
\sigma_{max}\,\approx\,\frac{2}{9}\frac{AB_e^3D_2}{\phi}\frac{q_0}{(L_{sat})^2},
\label{sigma_max}
\end{equation}

\begin{equation}
c_{max}\,\approx\,\frac{AB_eD_2}{\phi}\frac{q_0}{L_{sat}}.
\label{c_max}
\end{equation}

\section{Stability results}
\label{section:results}

The growth rate and the celerity, as expressed in Eqs. \ref{eq:sigma} and \ref{eq:c}, respectively, are computed next using the known values of coefficients $A$ and $B$, whereas some other terms, such as $\phi$, $u_*$, $d$ and $B_g$, are varied in order to analyze their effects on stability. For the entire analysis, it was considered that $A=3.2$ and $B=0.3$ because these are typical values for aquatic ripples \cite{Charru_5}; further, $U_s$ was computed by the Schiller--Naumann correlation \cite{Clift}. With regard to $C_{sat}$, no experimental measurements have yet been reported for the aquatic case \cite{Charru_5}. Then, $C_{sat}=5$ was adopted on the basis of the aeolian case, whose constant of proportionality is $4.4$ \cite{Claudin_Andreotti}.

The values of $B_g$ and $u_*$ are directly linked to the effects of gravity and fluid flow, respectively, and so, they are used next as the parameters to be varied. For small variations in grain diameter and flow far from the threshold (i.e., $D_2\approx 1$), variations in $d$ do not affect $D_2$ and do not imply significant variations in the settling velocity $U_s$. Therefore, in this case, the values of $d$ are directly linked to relaxation effects. Most previous stability analyses were conducted in this way, although shear stresses are close to threshold values when bed load occurs in water. However, close to the threshold and within large ranges of $d$, variations in $d$ cause considerable variations in the settling velocity as well as in the threshold shear stress. In this case, $d$ is closed linked to effects of threshold shear stress.

This section presents the variation of all these parameters and analyzes the influences of gravity, fluid velocities, relaxation, compactness, and threshold on initial instabilities. Whenever possible, the obtained diagrams are plotted in dimensional form in order to directly compare the stability outputs with experimental data.

\subsection{Stability diagrams}

\begin{figure}
   \begin{minipage}[c]{.49\linewidth}
    \begin{center}
     \includegraphics[width=\linewidth]{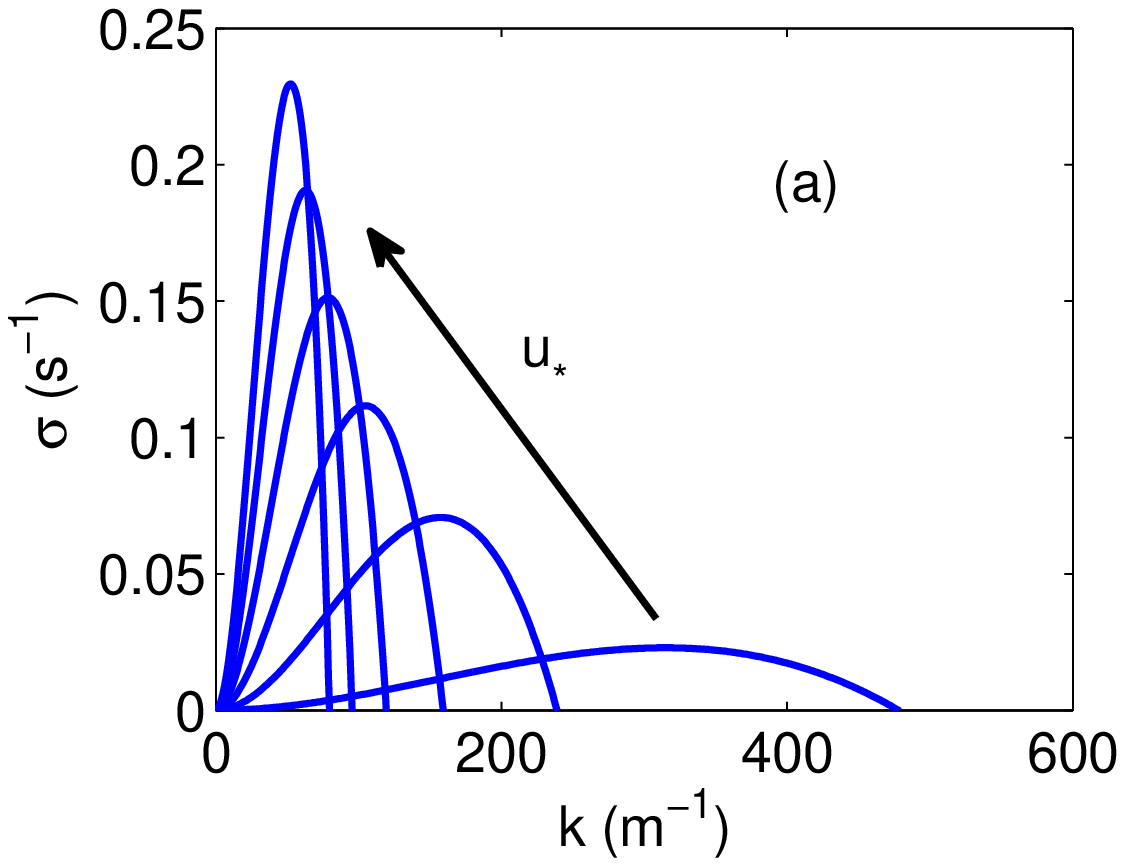}
    \end{center}
   \end{minipage} \hfill
   \begin{minipage}[c]{.49\linewidth}
    \begin{center}
      \includegraphics[width=\linewidth]{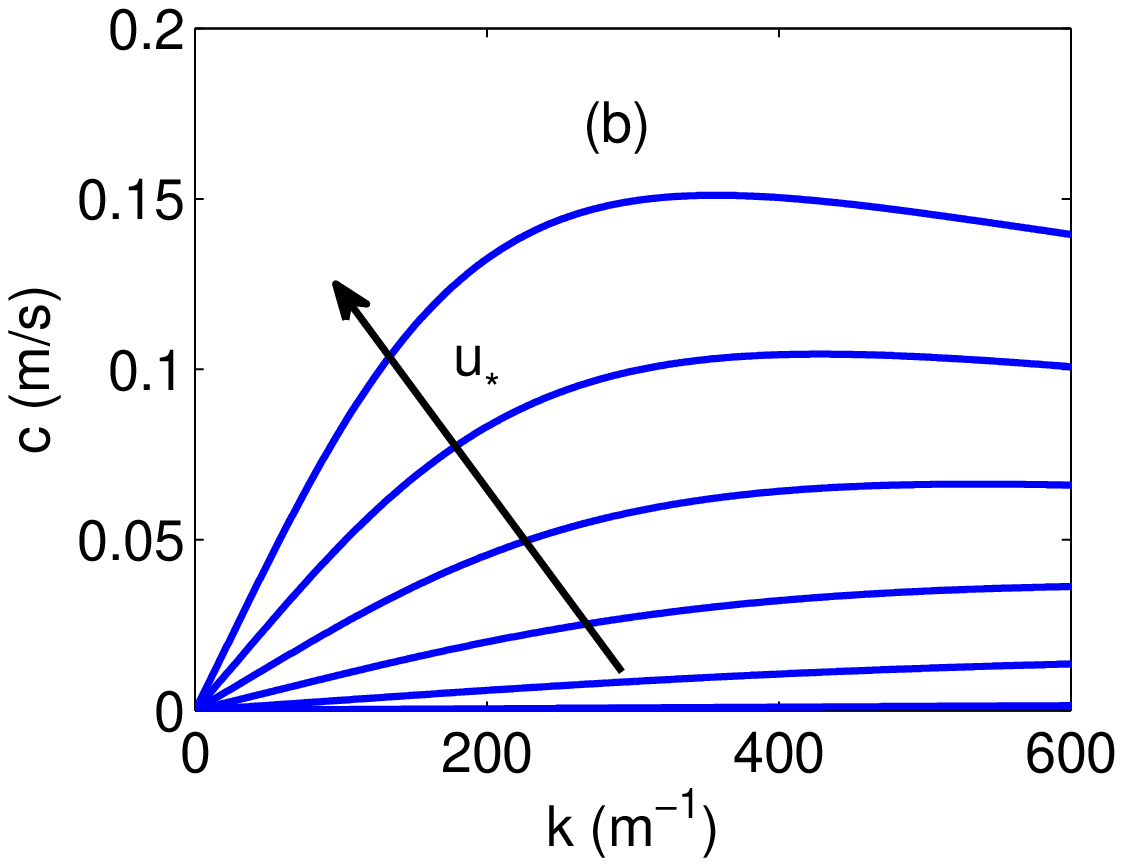}
    \end{center}
   \end{minipage}
\caption{(a) Growth rate $\sigma$ as a function of wavenumber $k$. (b) Phase velocity $c$ as a function of wavenumber $k$. The graphs are parameterized by the shear velocity $u_*$.}
\label{fig:sigma_c_varwith_u}
\end{figure}

\begin{figure}
   \begin{minipage}[c]{.49\linewidth}
    \begin{center}
     \includegraphics[width=\linewidth]{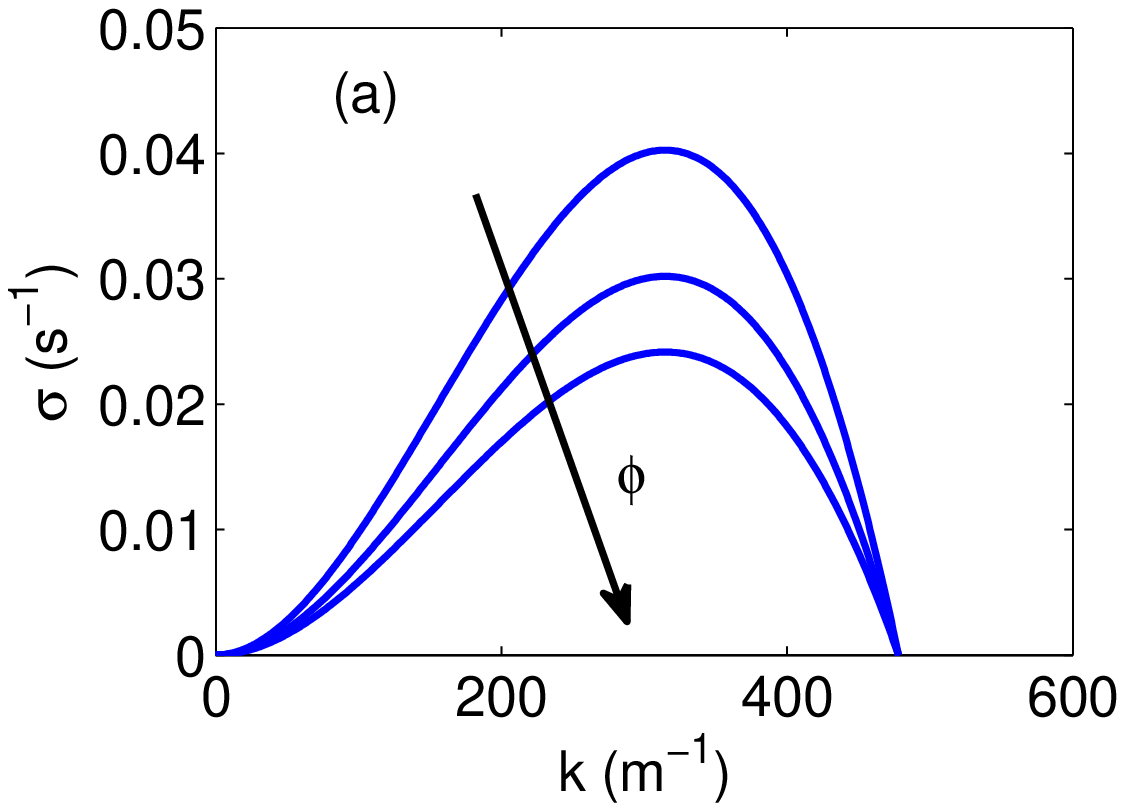}
    \end{center}
   \end{minipage} \hfill
   \begin{minipage}[c]{.49\linewidth}
    \begin{center}
      \includegraphics[width=\linewidth]{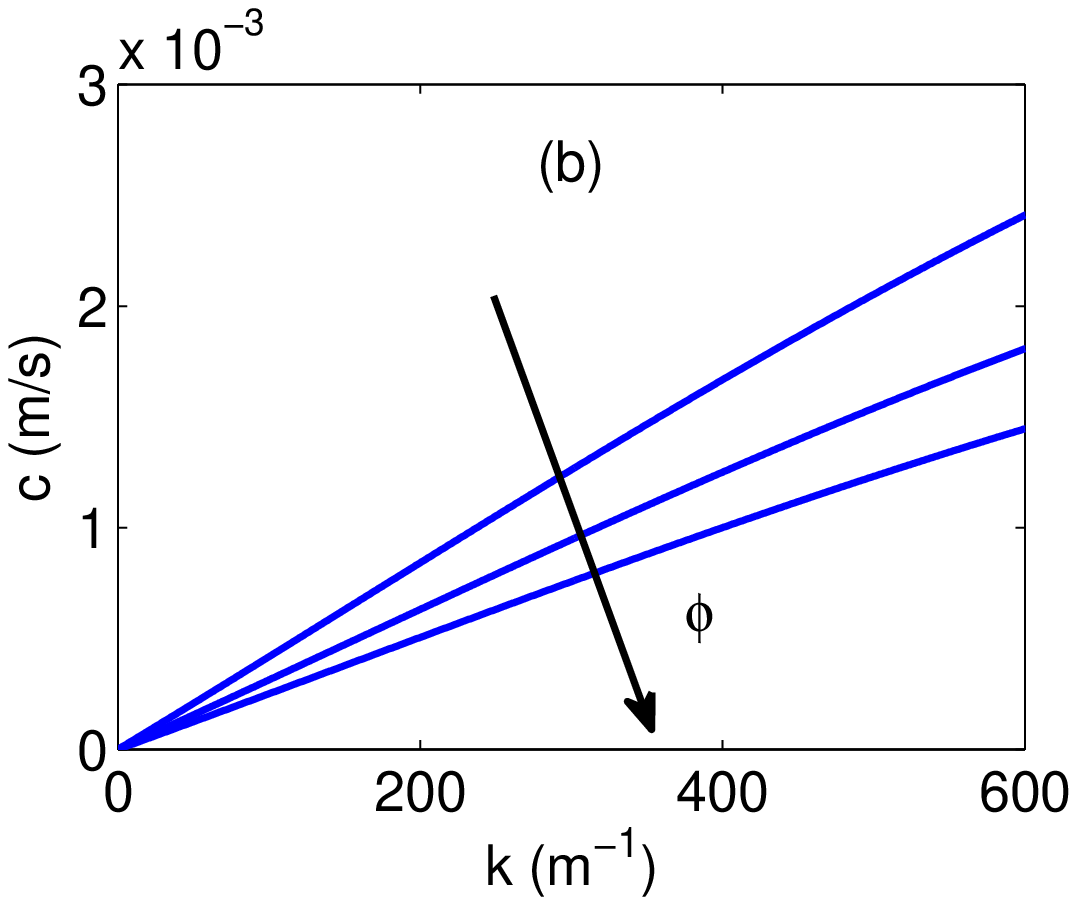}
    \end{center}
   \end{minipage}
\caption{(a) Growth rate $\sigma$ as a function of wavenumber $k$. (b) Phase velocity $c$ as a function of wavenumber $k$. The graphs are parameterized by the bed compactness $\phi$.}
\label{fig:sigma_c_varwith_phi}
\end{figure}

\begin{figure}
   \begin{minipage}[c]{.49\linewidth}
    \begin{center}
     \includegraphics[width=\linewidth]{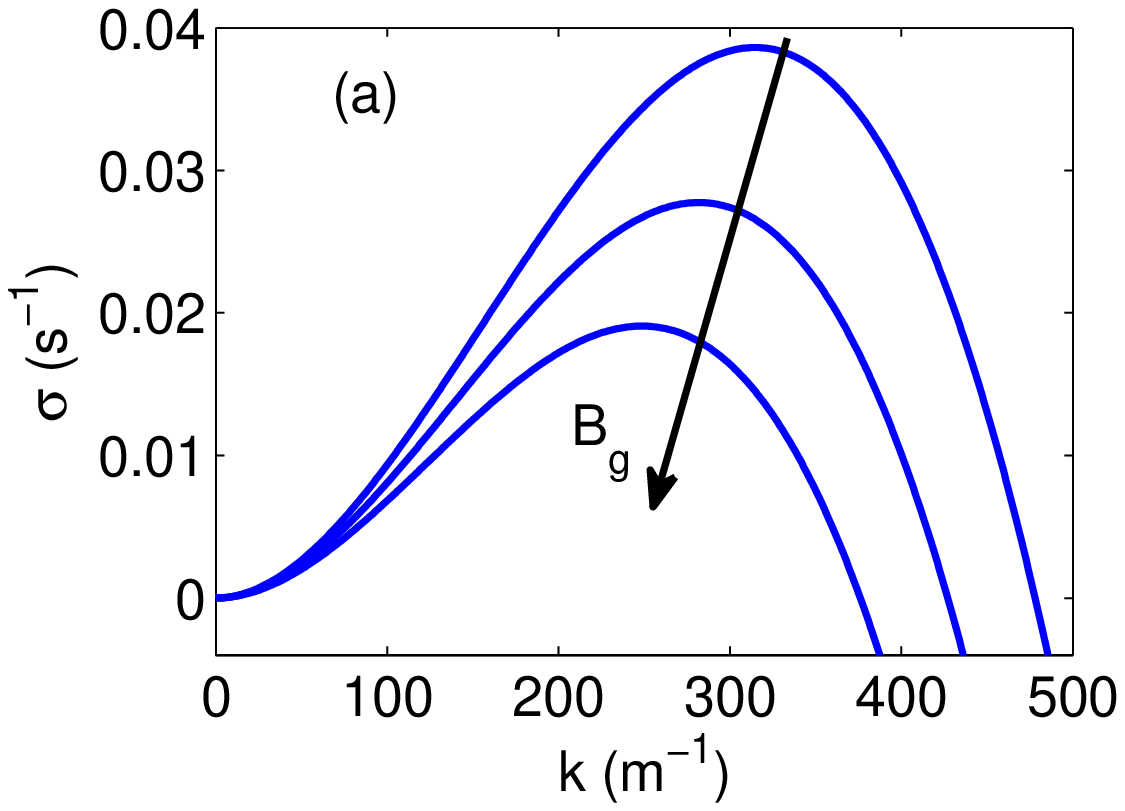}
    \end{center}
   \end{minipage} \hfill
   \begin{minipage}[c]{.49\linewidth}
    \begin{center}
      \includegraphics[width=\linewidth]{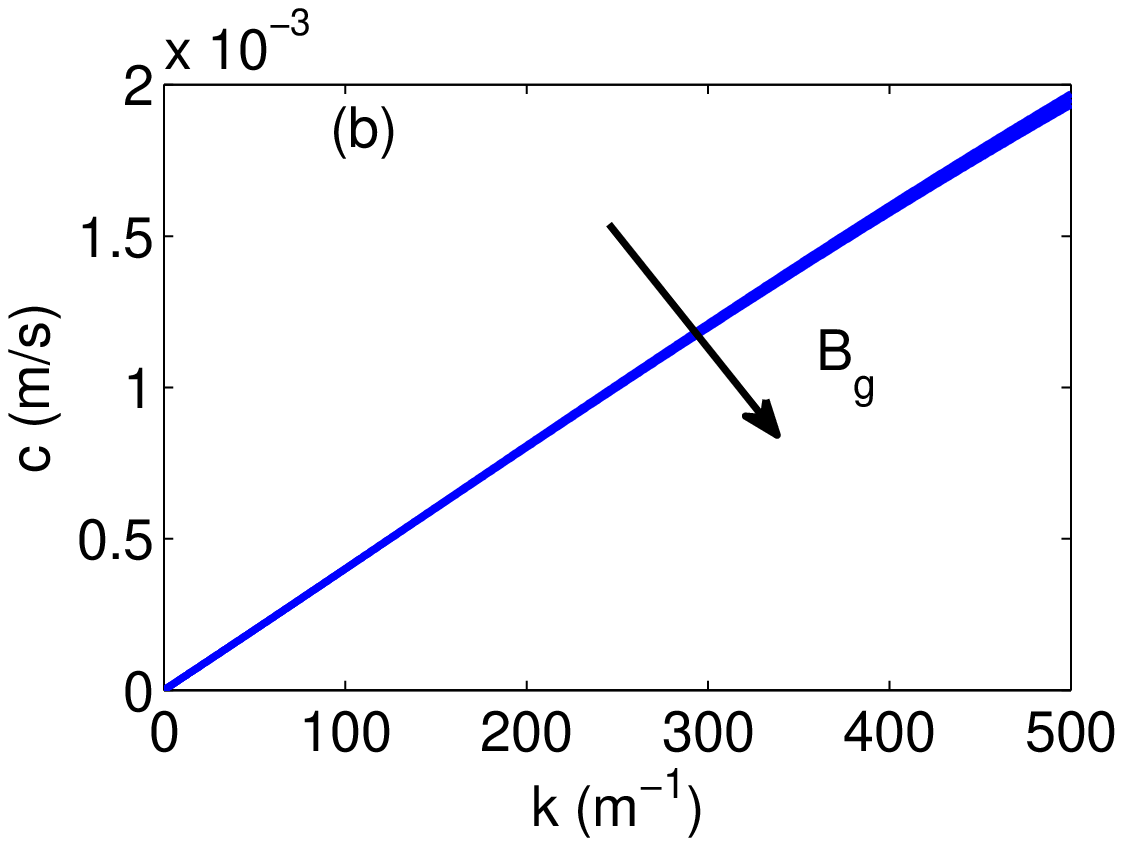}
    \end{center}
   \end{minipage}
\caption{(a) Growth rate $\sigma$ as a function of wavenumber $k$. (b) Phase velocity $c$ as a function of wavenumber $k$. The graphs are parameterized by the gravity coefficient $B_g$.}
\label{fig:sigma_c_varwith_bg}
\end{figure}

\begin{figure}
   \begin{minipage}[c]{.49\linewidth}
    \begin{center}
     \includegraphics[width=\linewidth]{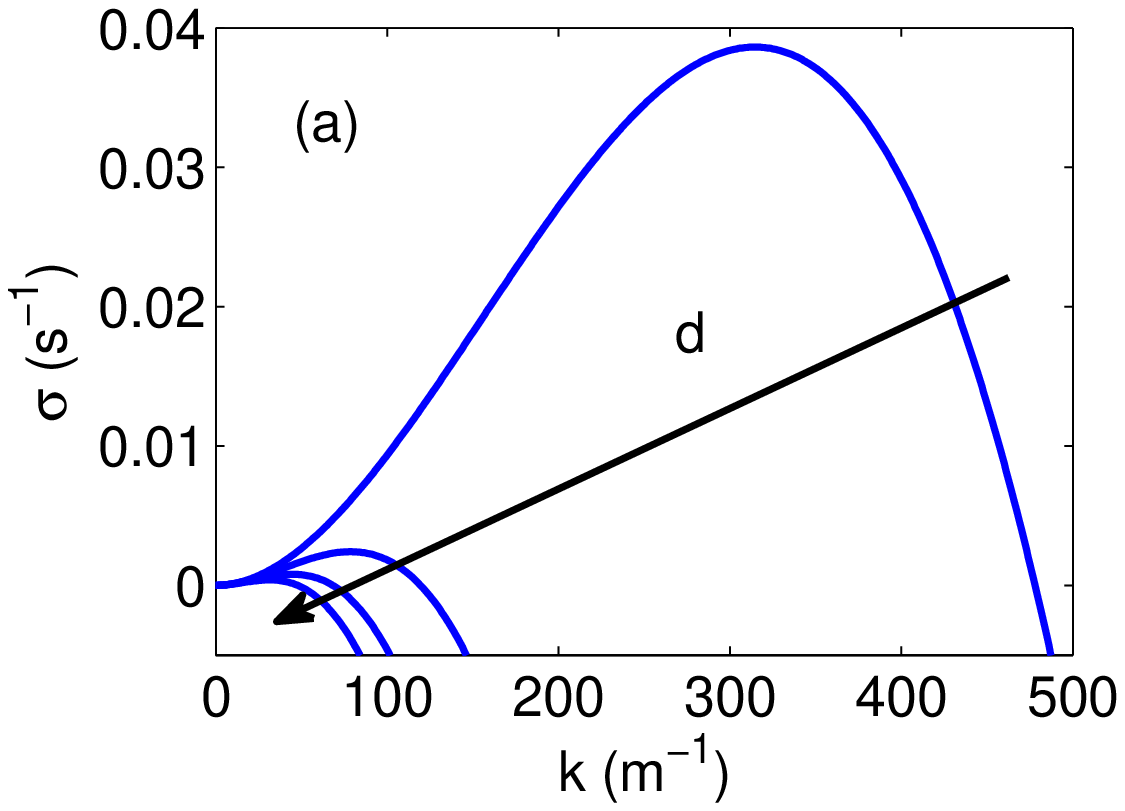}
    \end{center}
   \end{minipage} \hfill
   \begin{minipage}[c]{.49\linewidth}
    \begin{center}
      \includegraphics[width=\linewidth]{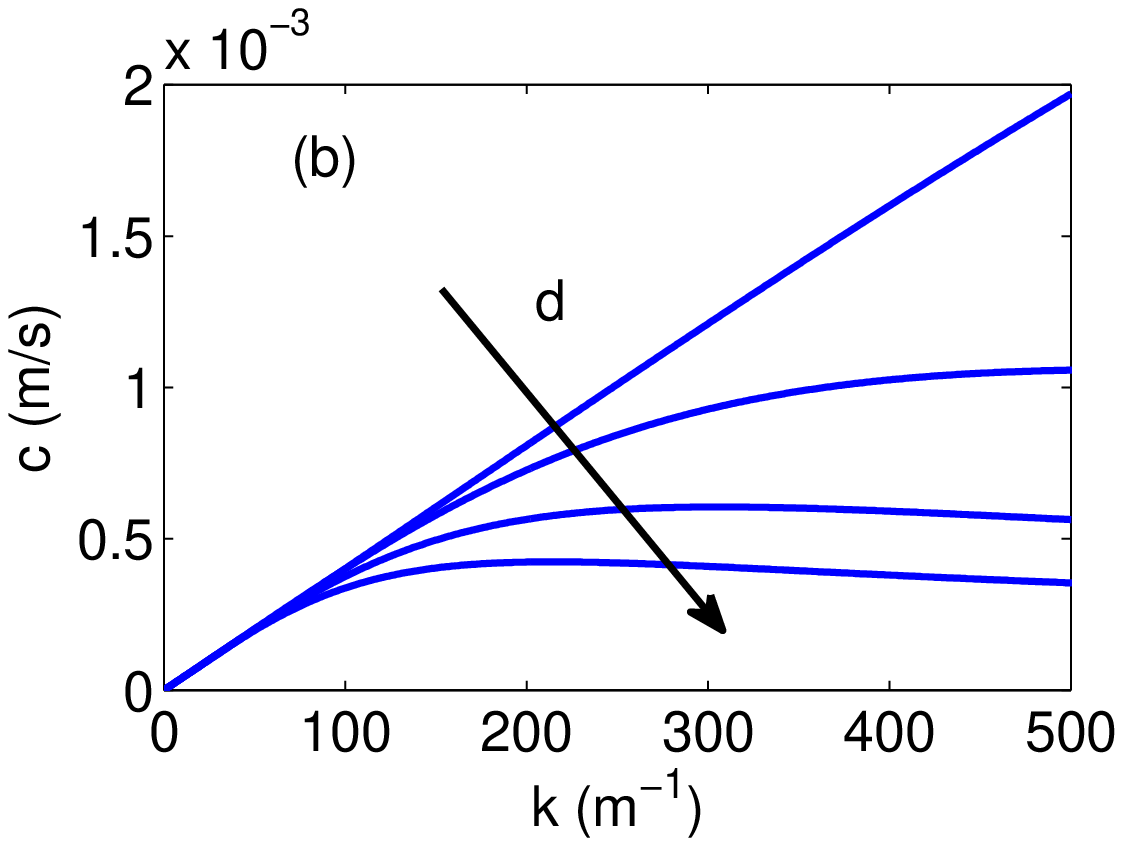}
    \end{center}
   \end{minipage}
\caption{(a) Growth rate $\sigma$ as a function of wavenumber $k$. (b) Phase velocity $c$ as a function of wavenumber $k$. The graphs are parameterized by varying only the grain diameter $d$.}
\label{fig:sigma_c_varwith_d}
\end{figure}

\begin{figure}
   \begin{minipage}[c]{.49\linewidth}
    \begin{center}
     \includegraphics[width=\linewidth]{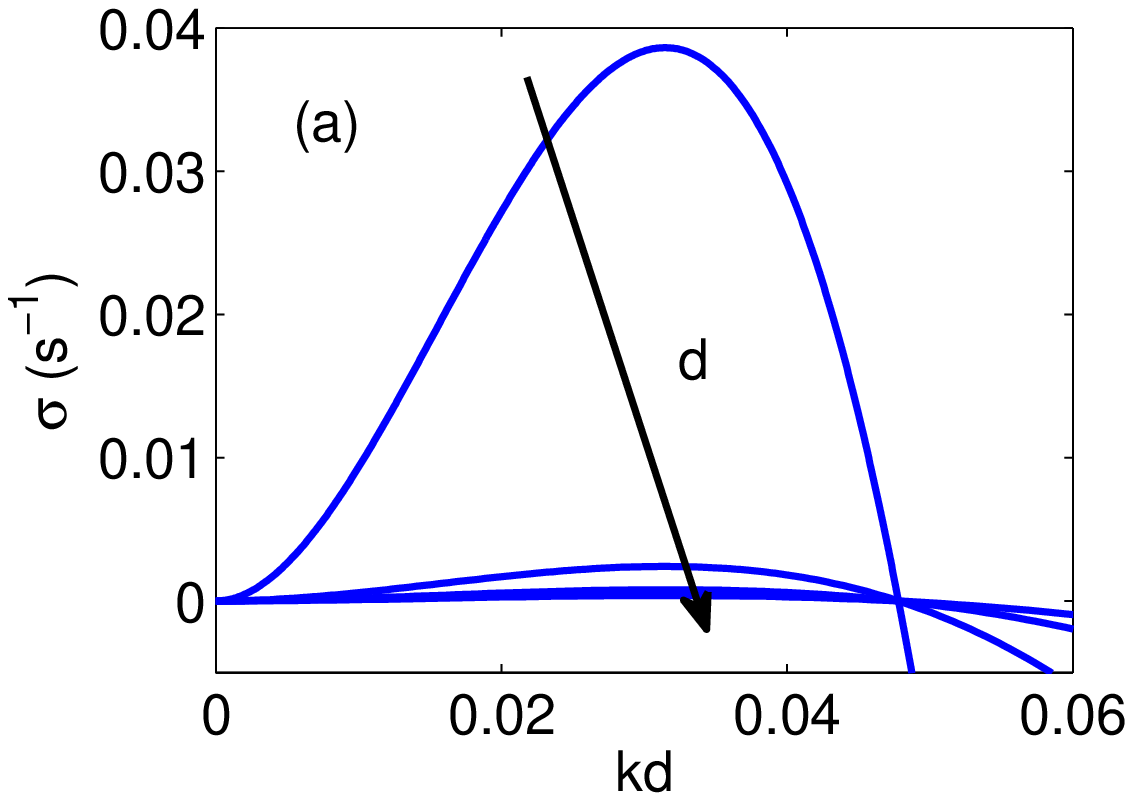}
    \end{center}
   \end{minipage} \hfill
   \begin{minipage}[c]{.49\linewidth}
    \begin{center}
      \includegraphics[width=\linewidth]{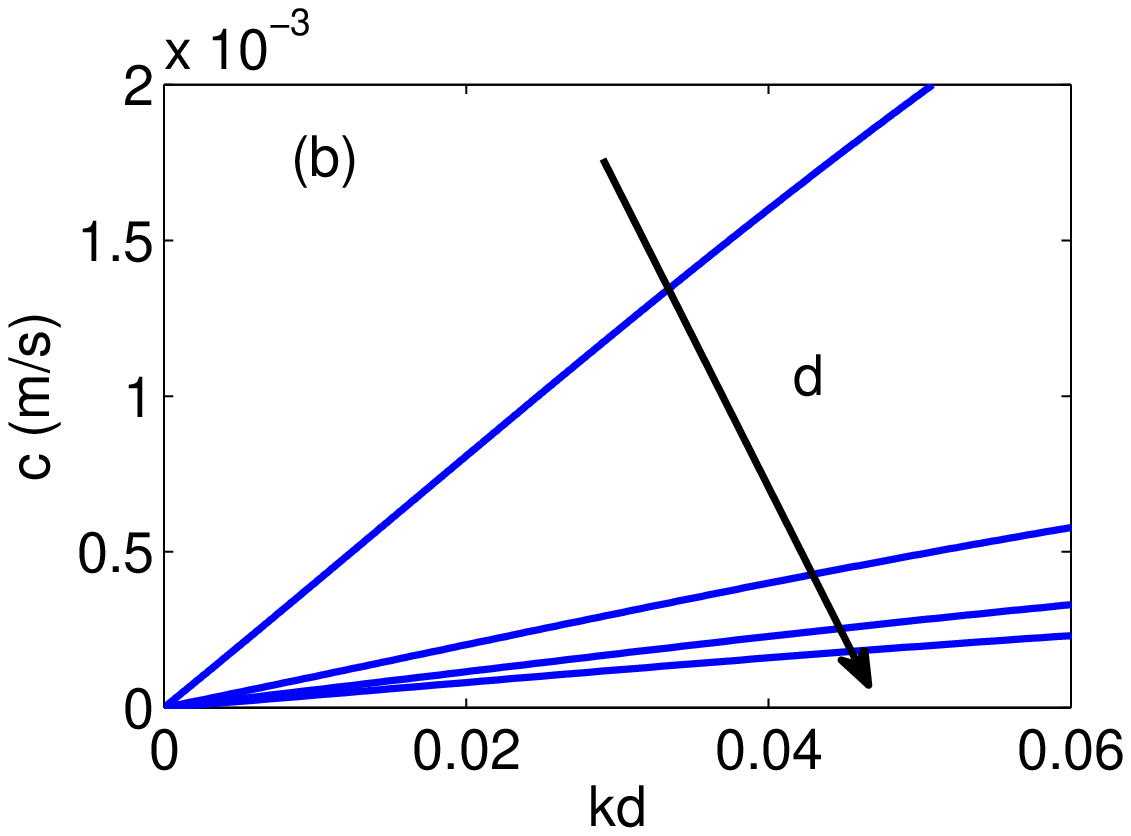}
    \end{center}
   \end{minipage}
\caption{(a) Growth rate $\sigma$ as a function of normalized wavenumber $kd$. (b) Phase velocity $c$ as a function of normalized wavenumber $kd$. The graphs are parameterized by varying only the grain diameter $d$.}
\label{fig:sigma_c_varwith_d_ad}
\end{figure}

\begin{figure}
   \begin{minipage}[c]{.49\linewidth}
    \begin{center}
     \includegraphics[width=\linewidth]{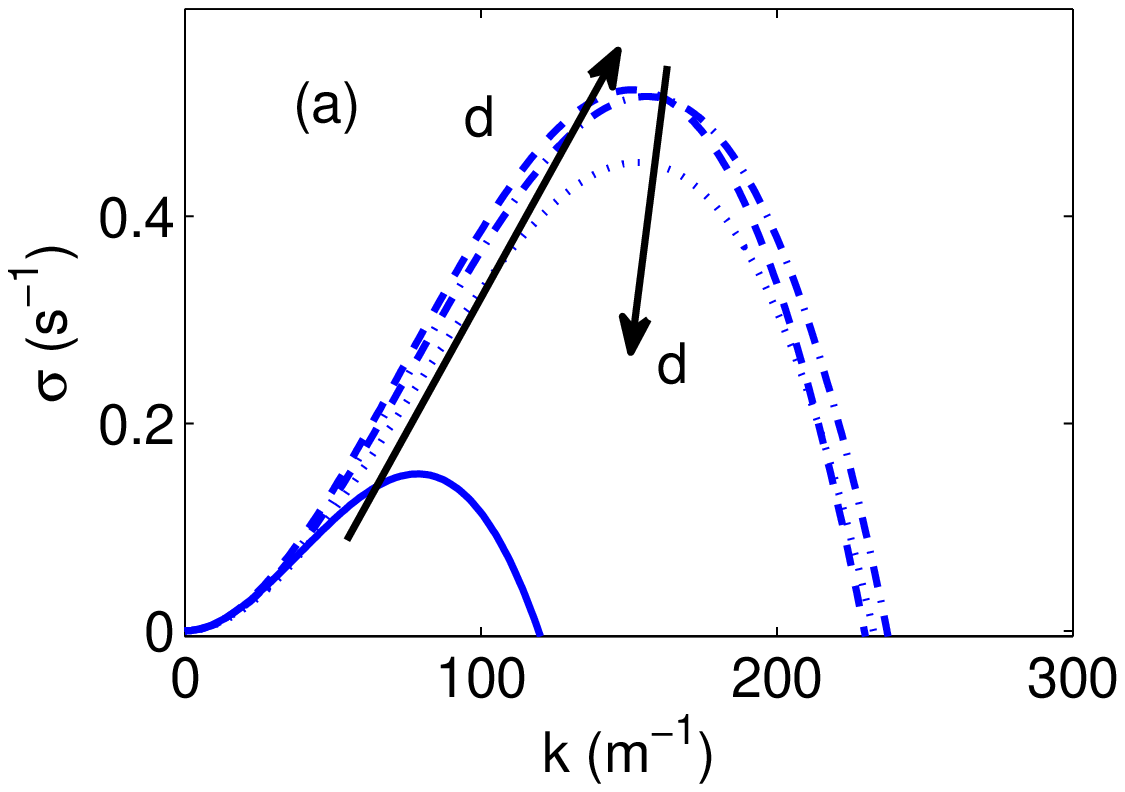}
    \end{center}
   \end{minipage} \hfill
   \begin{minipage}[c]{.49\linewidth}
    \begin{center}
      \includegraphics[width=\linewidth]{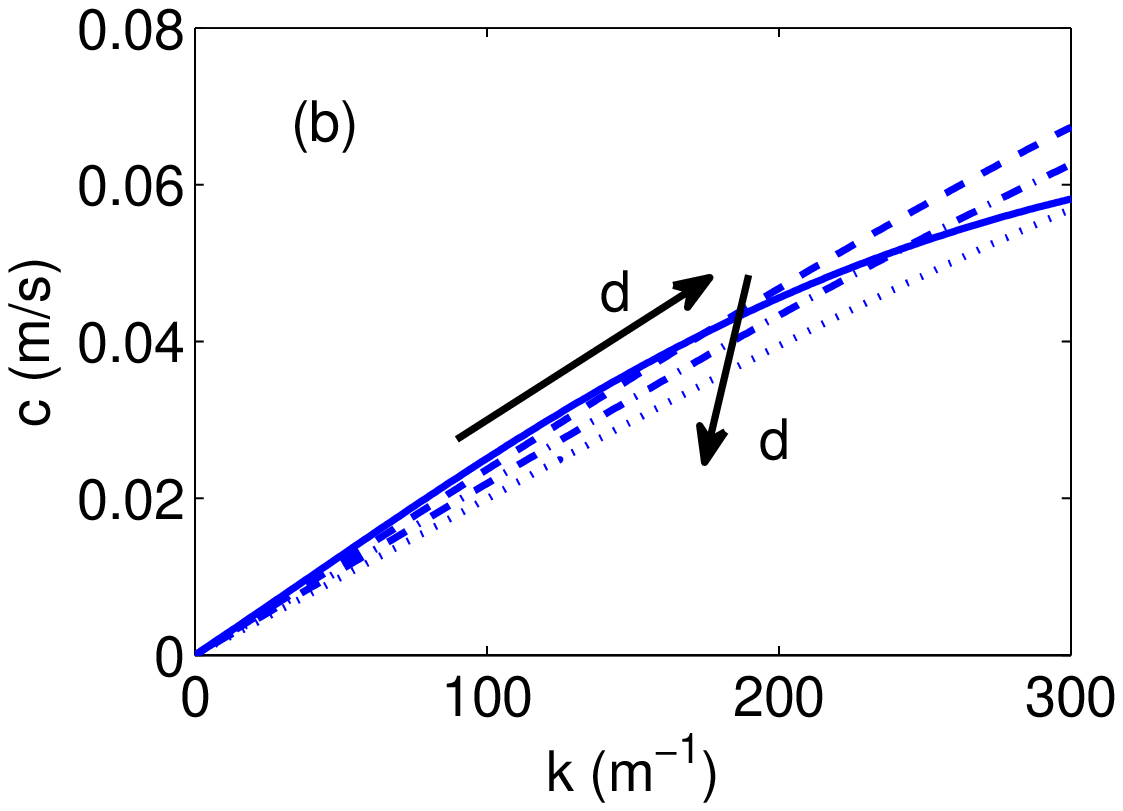}
    \end{center}
   \end{minipage}
\caption{(a) Growth rate $\sigma$ as a function of wavenumber $k$. (b) Phase velocity $c$ as a function of wavenumber $k$. The graphs are parameterized by the grain diameter $d$. The continuous, dashed, dashed-dotted, and dotted curves correspond to $d=0.1\,mm$, $d=0.4\,mm$, $d=0.7\,mm$, and $d=1\,mm$, respectively}
\label{fig:sigma_c_varwith_d2}
\end{figure}

Figures \ref{fig:sigma_c_varwith_u} to \ref{fig:sigma_c_varwith_d2} show the growth rates $\sigma$ and phase velocities $c$ as functions of the wavenumber $k$, parameterized by $u_*$, $\phi$, $B_g$, and $d$. The curves $\sigma (k)$ correspond to long-wave instabilities, where the long wavenumbers are always stable. Using $\sigma (k)$, it is possible to identify the unstable regions and the most unstable modes. For the corresponding wavenumbers, the celerity curves $c(k)$ give the celerity of each mode (usually, only the most unstable and the cutoff modes are of interest). Unless where deliberately varied or explicitly mentioned, the following values are fixed: $d=0.1\,mm$, $u_*=0.01\,m/s$, $\phi =0.6$, $B_g=0$, $\rho =1000\,kg/m^3$, $\rho_p =2500\,kg/m^3$, and $\mu =10^{-3}\,Pa.s$. These values are typical for bed-load transport of sand in water.

Figures \ref{fig:sigma_c_varwith_u}(a) and \ref{fig:sigma_c_varwith_u}(b), respectively, show the growth rate $\sigma$ and the phase velocity $c$ as functions of the wavenumber $k$. For obtaining these curves, the shear velocity was varied between $0.01\,m/s$ and $0.06\,m/s$ while the remaining terms were kept constant. The direction of growth in $u_*$ is shown also in these figures.

\begin{sloppypar}
For the most unstable mode, these curves agree with the variation obtained with the long-wave approximation. For the growth rate, celerity, and wavenumber, Eqs. \ref{k_max}, \ref{sigma_max}, and \ref{c_max} predict $k_{max}\,\sim\,u_*^{-1}$, $\sigma_{max}\,\sim\,u_*^2\left( u_*^2-u_{th}^2 \right)^{-1} \left( u_*^2-u_{th}^2 \right)^{3/2} u_*^{-2}$, and $c_{max}\,\sim\,u_*^2\left( u_*^2-u_{th}^2 \right)^{-1} \left( u_*^2-u_{th}^2 \right)^{3/2} u_*^{-1}$ as given by the wavenumbers corresponding to the maxima in Fig. \ref{fig:sigma_c_varwith_u}(a).
\end{sloppypar}

Far from the threshold, $u_*\gg u_{th}$ and $\sigma_{max}\,\sim\,u_*$, $c_{max}\,\sim\,u_*^2$, and $k_{max}\,\sim\,u_*^{-1}$, just as predicted by previous analyses where the threshold effects were neglected (e.g., \cite{Franklin_4}). This behavior approaches situations where shear velocities are high for the corresponding grain diameters (or the diameters are small for the corresponding shear velocities), for instance, curves for high $u_*$ (the three leftmost curves) in Fig. \ref{fig:sigma_c_varwith_u}(a). However, the behavior is different when the shear velocities are moderate for the corresponding grain diameters, as is the case of the curves for low $u_*$ (the three rightmost curves) in Fig. \ref{fig:sigma_c_varwith_u}(a).

In liquids, bed load is never too far from threshold. Therefore, threshold effects cannot be neglected and Eqs. \ref{eq:sigma}--\ref{c_max} are then used.

Figures \ref{fig:sigma_c_varwith_phi}(a) and \ref{fig:sigma_c_varwith_phi}(b), respectively, show the growth rate $\sigma$ and the phase velocity $c$ as functions of the wavenumber $k$. For obtaining these curves, the bed compactness was varied between $0.6$ and $1$, while the remaining terms were kept constant. The direction of growth in $\phi$ is also shown in these figures.

As predicted by Eqs. \ref{k_max}--\ref{c_max}, the behaviors of the growth rate and the celerity of the most unstable mode are 
$\sigma_{max}\,\sim\,\phi^{-1}$ and $c_{max}\,\sim\,\phi^{-1}$, respectively, whereas the wavenumber and wavelength do not vary with $\phi$.

In practical situations, the bed compactness is $\phi\approx 0.6$ and the dispersion around this value is small. As the growth rate and the celerity vary as a unit power of $1/\phi$, the influence of the bed compactness on the growth rate and celerity may be neglected.

Figures \ref{fig:sigma_c_varwith_bg}(a) and \ref{fig:sigma_c_varwith_bg}(b), respectively, show the growth rate $\sigma$ and the phase velocity $c$ as functions of the wavenumber $k$. For obtaining these curves, the gravity coefficient $B_g$ was varied between $0$ and $0.02$ and the remaining terms were kept constant. The direction of growth in $B_g$ is shown in these figures.

Figure \ref{fig:sigma_c_varwith_bg}(a) shows that $\sigma_{max}$ decreases as $B_g$ increases. This agrees with Eq. \ref{sigma_max}, which predicts $\sigma_{max}\,\sim\,\left( B-B_g/A\right)^3$. However, the value of $B_g$ has not been determined experimentally and can only be estimated by dimensional analysis. In the case of water, it is estimated to be $ord(0.01)$ \cite{Charru_3}, where the term $ord$ denotes \textit{order of magnitude}. As $B=ord(0.1)$ and $A=ord(1)$, the effects of $B_g$ on $\sigma$ are not very pronounced, as is expected for water given the low relative weight of grains and the lubricating effect of water. Figure \ref{fig:sigma_c_varwith_bg}(a) shows that the wavenumber decreases linearly with the growth of $B_g$, in accordance with Eq. \ref{k_max}, which predicts $k_{max}\,\sim\,-B_g$. Then, gravity effects tend to increase the most unstable wavelength, and the same occurs for the cutoff wavelength. Further, Fig. \ref{fig:sigma_c_varwith_bg}(b) shows that the celerity remains almost unchanged by $B_g$, which (considering the orders of magnitude of $A$, $B$, and $B_g$) agrees with Eq. \ref{c_max}. Finally, gravity effects in water are smaller than other effects, and given the large uncertainties in estimating $B_g$, they may be neglected.

Figures \ref{fig:sigma_c_varwith_d}(a) and \ref{fig:sigma_c_varwith_d}(b), respectively, show the growth rate $\sigma$ and the phase velocity $c$ as functions of the wavenumber $k$. For obtaining these curves, the grain diameter $d$ was varied between $0.1\,mm$ and $1\,mm$, while the remaining terms except for $L_{sat}$ were kept constant. The direction of growth in $d$ is also shown in these figures.

The main objective of these figures is to understand the effects of relaxation on initial instabilities. In order to isolate these effects, $D_2$ and the settling velocity $U_s$ were kept constant but $d$ was varied. Physically, this corresponds to small variations in grain diameter and flows far from the threshold, which is not really expected for liquids, but is a rough approximation for high shear velocities. In obtaining Figs. \ref{fig:sigma_c_varwith_d}(a) and \ref{fig:sigma_c_varwith_d}(b), $D_2$ was fixed at unity and $U_s$ was fixed to the value corresponding to $d=0.1\,mm$.

Figure \ref{fig:sigma_c_varwith_d}(a) shows a strong decrease in both the growth rate and the wavenumber with increasing grain diameter. This agrees with Eqs. \ref{k_max} and \ref{sigma_max}, which predict (for variations in only $d$) $k_{max}\,\sim\,d^{-1}$ and $\sigma_{max}\,\sim\,d^{-2}$. Figure \ref{fig:sigma_c_varwith_d}(a) also shows that the cutoff wavenumber varies with $d^{-1}$. This strong behavior should not be expected in the case of liquids: to the best of the author's knowledge, such strong variation of $\sigma_{max}$ with $d$ has not yet been reported.

Concerning the celerities, it is difficult to analyze  Fig. \ref{fig:sigma_c_varwith_d}(b) because the most unstable wavenumber is different for each $d$. In order to compare the celerities corresponding to $k_{max}$ for each diameter, the abscissas of the graphs in Fig. \ref{fig:sigma_c_varwith_d} were normalized by $d$ in Fig. \ref{fig:sigma_c_varwith_d_ad}. Figure \ref{fig:sigma_c_varwith_d_ad}(a) shows that $k_{max}d\,\approx\,0.03$, and Fig. \ref{fig:sigma_c_varwith_d_ad}(b) shows a decrease in the celerity with increasing grain diameter. This agrees with Eq. \ref{c_max}, which predicts $c_{max}\,\sim\,d^{-1}$ in this case. However, this dependency is contrary to experimental observations, in which the celerity increases by a small amount with increasing grain diameter \cite{Franklin_3}.

The large differences between the linear analysis and the experimental results indicate that stability analyses based solely on variations of grain diameter, without taking into account the threshold and settling velocity variations, are unsuitable for bed load in liquids.

Figures \ref{fig:sigma_c_varwith_d2}(a) and \ref{fig:sigma_c_varwith_d2}(b), respectively, show the growth rate $\sigma$ and the phase velocity $c$ as functions of the wavenumber $k$. For obtaining these curves, the grain diameter $d$ was varied between $0.1\,mm$ and $1\,mm$, while the remaining terms that do not depend on $d$ were kept constant. The directions of growth in $d$ are also shown in these figures. The continuous, dashed, dashed-dotted, and dotted curves correspond to $d=0.1\,mm$, $d=0.4\,mm$, $d=0.7\,mm$, and $d=1\,mm$, respectively. In Fig. \ref{fig:sigma_c_varwith_d2}, the terms depending on $d$, i.e., $L_{sat}$, $D_2$, and $U_s$, were varied accordingly. This means that the graphs in this figure are suitable for large variations of $d$ and for flow conditions close to the threshold; therefore, these graphs can highlight the threshold effects.

Figure \ref{fig:sigma_c_varwith_d2}(a) shows that the most unstable mode varies non monotonically with $d$: the wavenumber first increases and then decreases. This variation is weaker than that in the case of Figs. \ref{fig:sigma_c_varwith_d} and \ref{fig:sigma_c_varwith_d_ad}, and, in principle, does not agree with experimental observations that the wavelength of aquatic ripples increases with $d$ \cite{Charru_5, Coleman_1}. However, measured aquatic ripples correspond to different fluid flow conditions, where $u_*$ is usually larger for larger diameters. In this sense, $u_*$ shall be varied when comparing the stability analysis with experimental data. This is done as shown in Fig. \ref{fig:d2_varwith_u}, which is presented in the next subsection. With regard to the celerity, Fig. \ref{fig:sigma_c_varwith_d2}(b) shows that $c_{max}$ also varies non monotonically with $d$, and for the same reason as that given for $\sigma$, comparison with experimental data must take $u_*$ into account.

Variations in $d$ accompanied by the respective variations of $L_{sat}$, $D_2$, and $U_s$, which correspond to combined threshold and relaxation effects, are weaker than variations in only $d$, which correspond to isolated relaxation effects, and are closest to experimental measurements (presented next). When considering different values of $u_*$, this combined threshold-relaxation analysis is suitable for predicting the formation of ripples.

\begin{figure}
   \begin{minipage}[c]{.49\linewidth}
    \begin{center}
     \includegraphics[width=\linewidth]{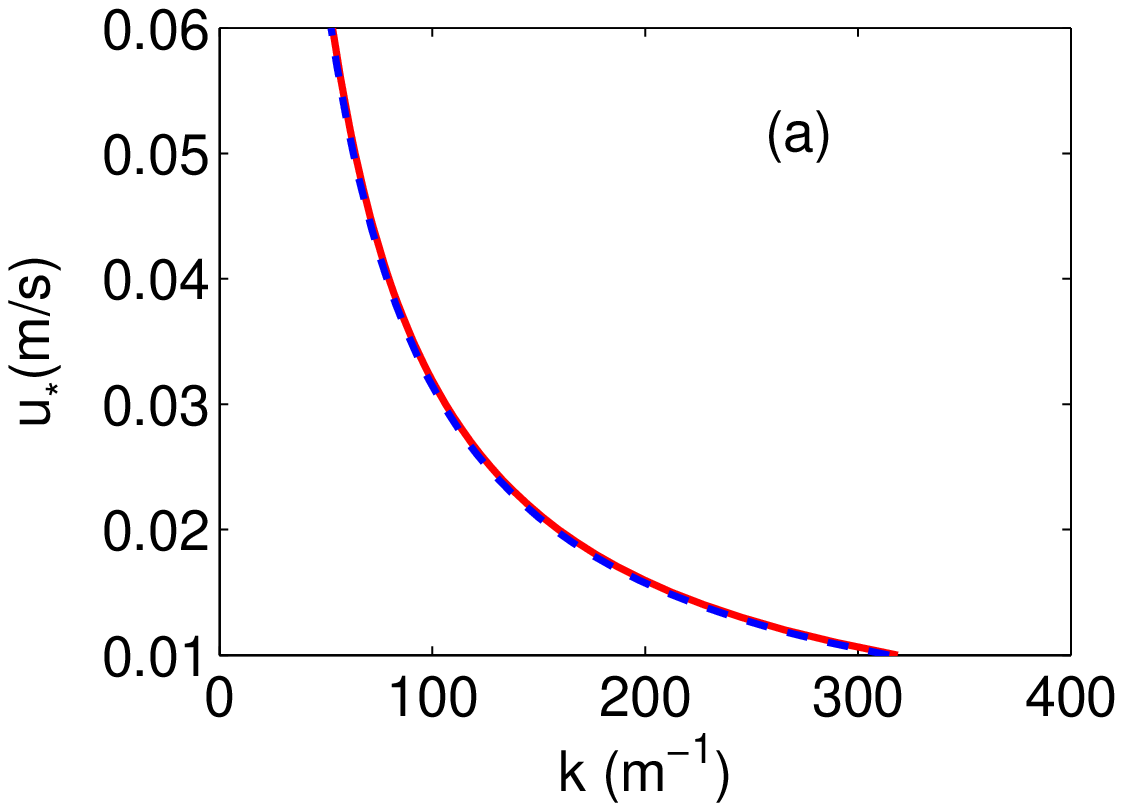}
    \end{center}
   \end{minipage} \hfill
   \begin{minipage}[c]{.49\linewidth}
    \begin{center}
      \includegraphics[width=\linewidth]{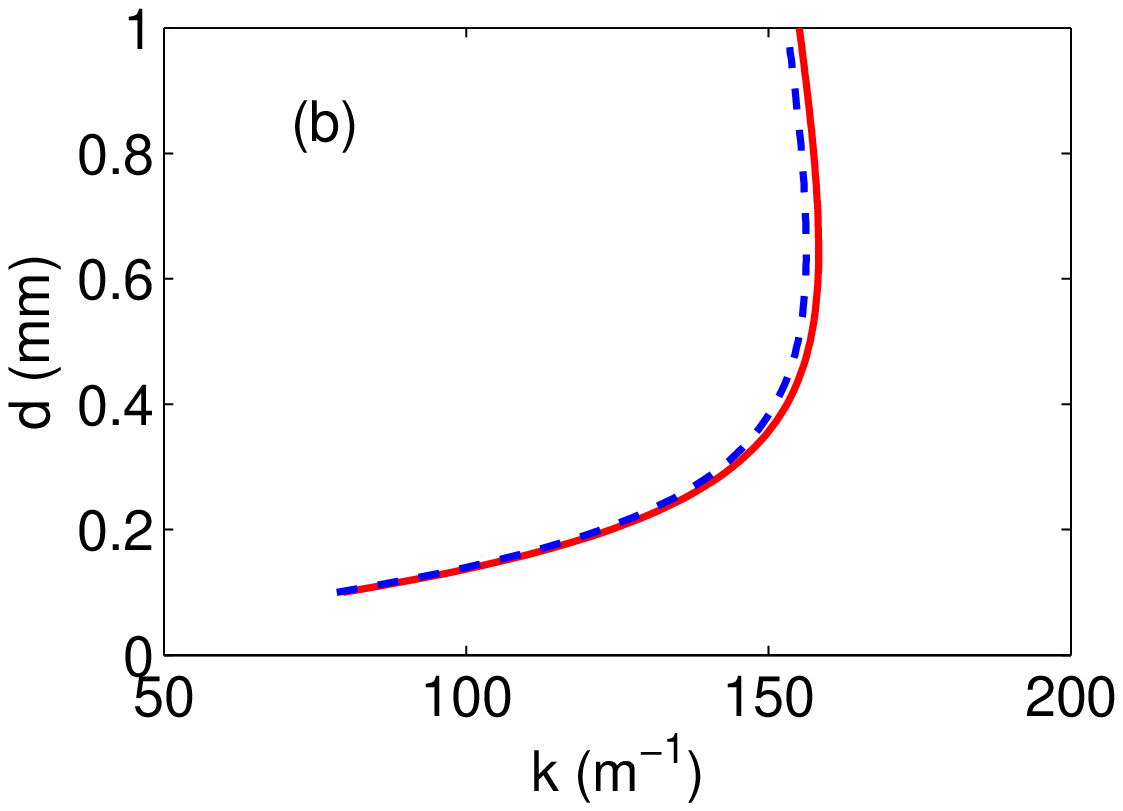}
    \end{center}
   \end{minipage}
\caption{(a) Curves of the most unstable modes in terms of $u_*$. (b) Curves of the most unstable modes in terms of $d$, where all the parameters affected by $d$ were varied accordingly. The continuous curves correspond to the long-wave approximation (Eq. \ref{sigma_max}) and the dashed curves correspond to the maxima of the complete solution (Eq. \ref{eq:sigma}).}
\label{fig:most_unstable_modes}
\end{figure}

Finally, Fig. \ref{fig:most_unstable_modes} compares the maxima of $\sigma(k)$, given by Eq. \ref{eq:sigma} and plotted in Figs. \ref{fig:sigma_c_varwith_u}(a) and \ref{fig:sigma_c_varwith_d2}(a), with the long-wave approximation given by Eq. \ref{sigma_max}. The two curves are almost superposed, revealing that the long-wave approximation is a good one and justifying \textit{a posteriori} its use.

\subsection{Marginal stability}

\begin{figure}
   \begin{minipage}[c]{.49\linewidth}
    \begin{center}
     \includegraphics[width=\linewidth]{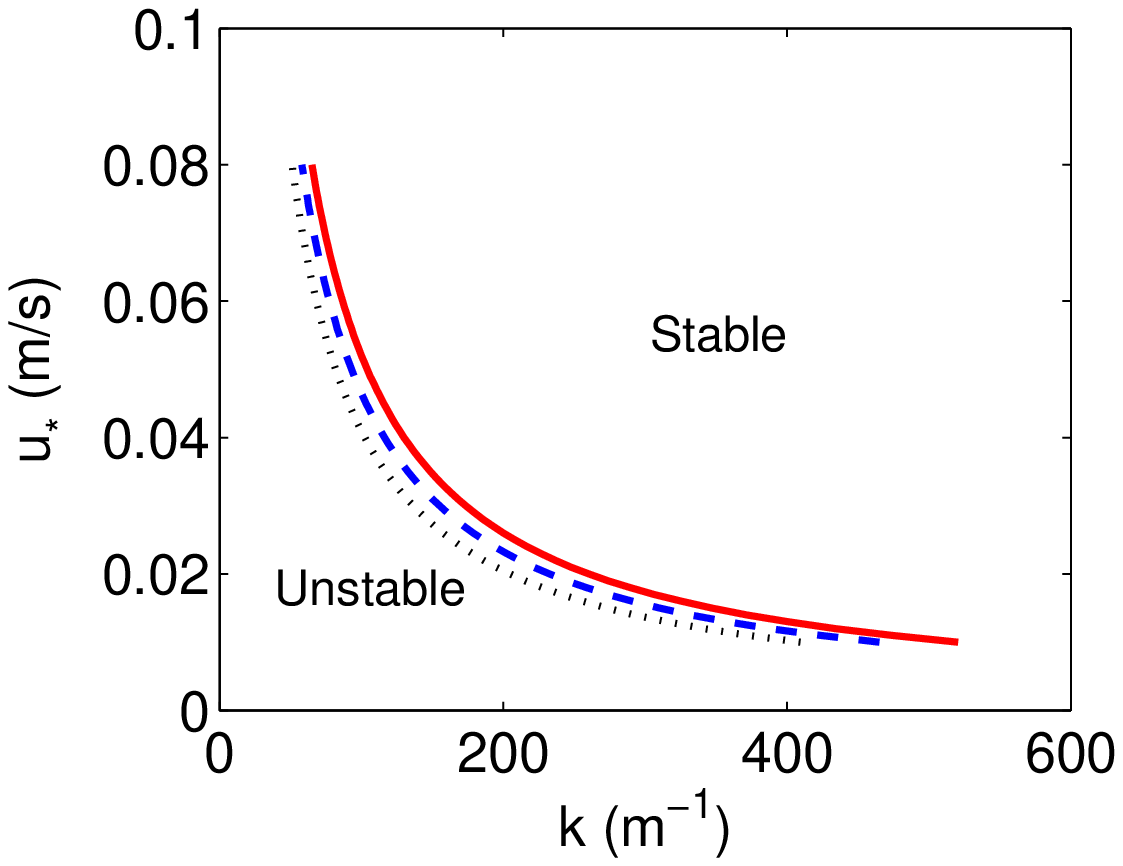}
    \end{center}
   \end{minipage} \hfill
   \begin{minipage}[c]{.49\linewidth}
    \begin{center}
      \includegraphics[width=\linewidth]{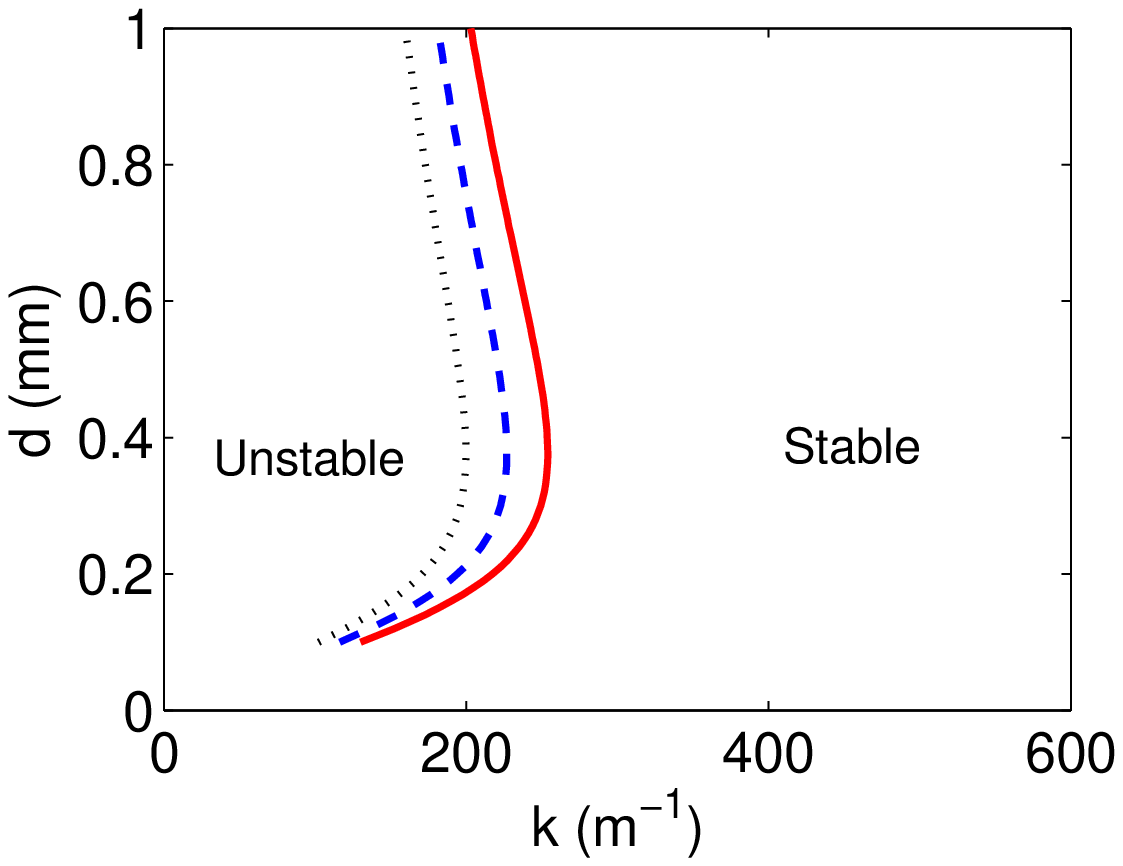}
    \end{center}
   \end{minipage}
\caption{(a) Curves of marginal stability in terms of $u_*$ for different values of $B_g$. (b) Curves of marginal stability in terms of $d$ for different values of $B_g$, where all the parameters affected by $d$ were also varied. The continuous, dashed, and dotted curves correspond to $B_g=0$, $B_g=0.01$, and $B_g=0.02$, respectively.}
\label{fig:marg_varwith_bg}
\end{figure}

Figure \ref{fig:marg_varwith_bg}(a) shows curves of marginal stability in terms of $u_*$ (fluid flow effects) for different values of $B_g$ (gravity effects), and Fig. \ref{fig:marg_varwith_bg}(b) shows the curves of marginal stability in terms of $d$ for different values of $B_g$, where all the parameters affected by $d$ were varied accordingly (relaxation effects and threshold effects were combined). Figure \ref{fig:marg_varwith_bg} shows that gravity effects on the cutoff wavenumber are smaller than the effects of the threshold, relaxation, and fluid flow. Given the large uncertainties in estimating $B_g$ and its relatively small effects, gravity effects may be neglected in the case of liquids.

\begin{figure}
   \begin{minipage}[c]{\columnwidth}
    \begin{center}
     \includegraphics[width=.60\linewidth]{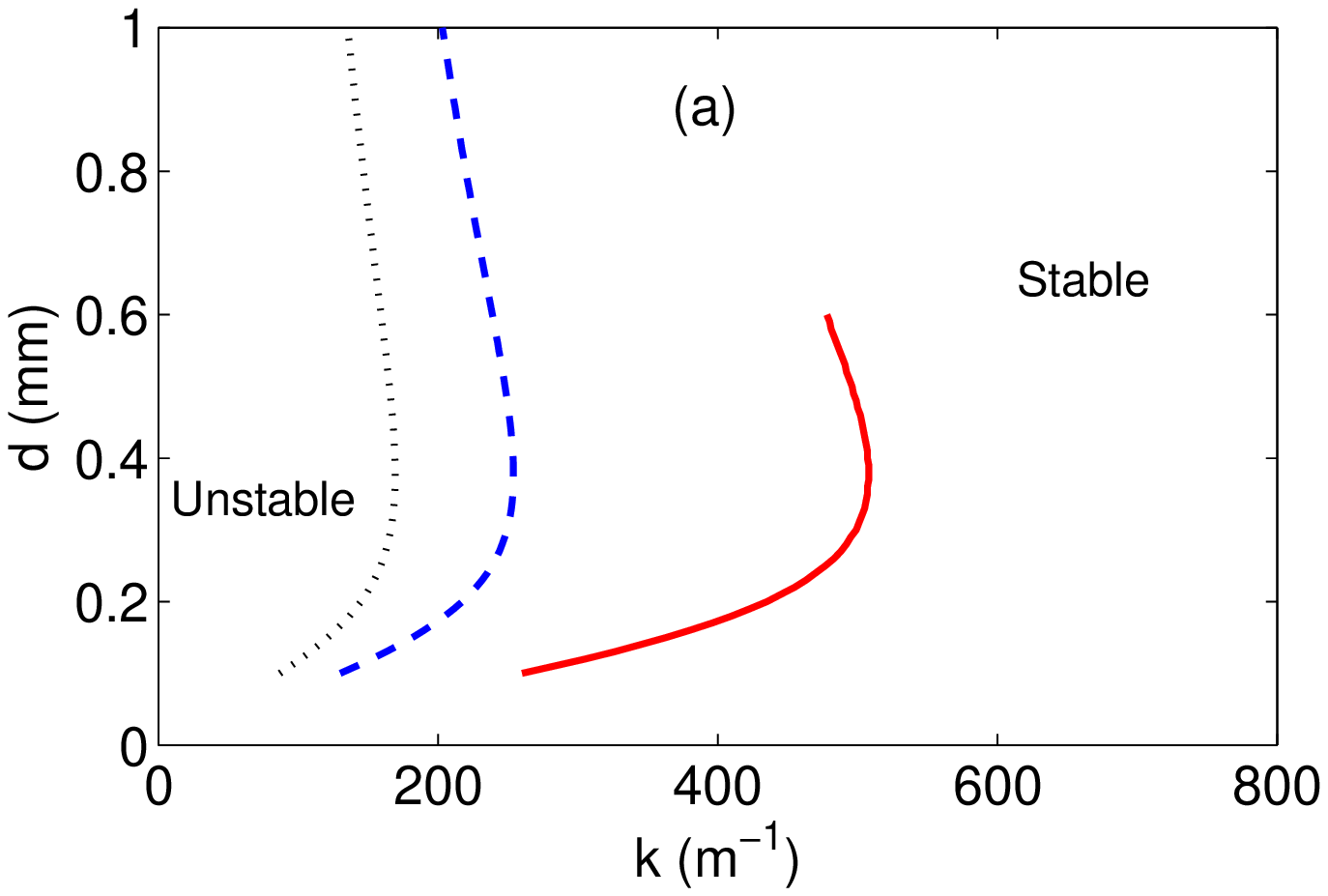}
    \end{center}
   \end{minipage} \hfill
   \begin{minipage}[c]{\columnwidth}
    \begin{center}
      \includegraphics[width=.60\linewidth]{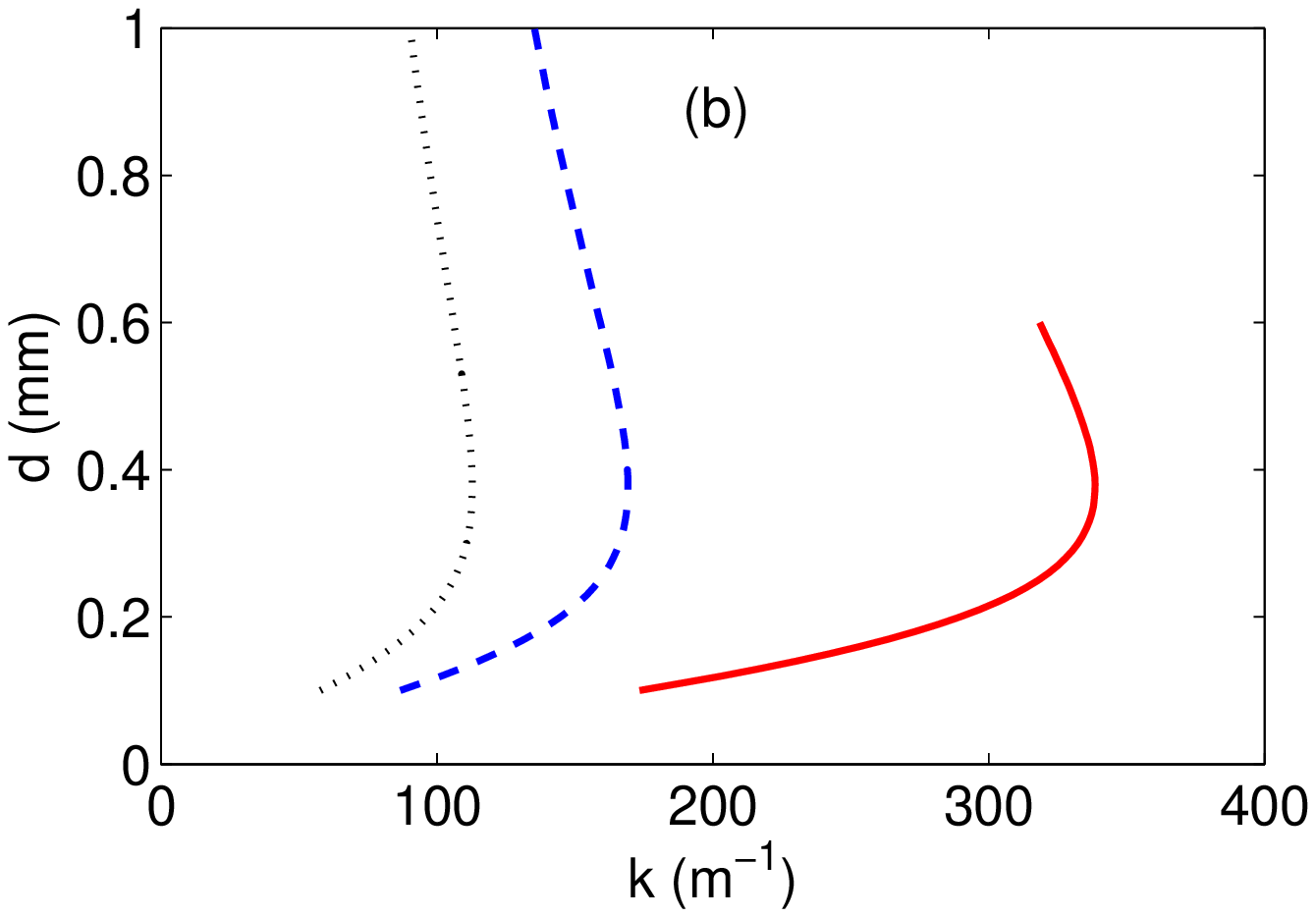}
    \end{center}
   \end{minipage}
\caption{(a) Curves of marginal stability in terms of $d$ for different values of $u_*$. (b) Curves of the most unstable modes in terms of $d$ for different values of $u_*$. In these curves, all the parameters affected by $d$ were also varied. The continuous, dashed, and dotted curves correspond to $u_*=0.02\,m/s$, $u_*=0.04\,m/s$, and $u_*=0.06\,m/s$, respectively.}
\label{fig:d2_varwith_u}
\end{figure}

Figure \ref{fig:d2_varwith_u}(a) shows curves of marginal stability in terms of $d$ for different values of $u_*$, and Fig. \ref{fig:d2_varwith_u}(b) shows the curves of the most unstable modes in terms of $d$ for different values of $u_*$. In these curves, all the parameters affected by $d$ were varied accordingly, and so, the curves correspond to the marginal stability of the combined threshold and relaxation effects, parameterized by fluid flow effects. The continuous, dashed, and dotted curves correspond to $u_*=0.02\,m/s$, $u_*=0.04\,m/s$, and $u_*=0.06\,m/s$, respectively. Figure \ref{fig:d2_varwith_u} shows that the cutoff and the most unstable wavenumbers vary strongly with the shear velocity. This means that fluid flow effects on both the marginal stability and the most unstable mode are strong and must be taken into account. 

In summary, the linear stability analysis predicts that fluid flow, threshold, and relaxation effects are strong and must be taken into account when predicting the formation of aquatic ripples. In order to verify this prediction, the linear stability analysis is compared with some experimental results in the next subsection.

\subsection{Comparison with experimental results}

Next, the results of the above-presented analysis are compared with published experimental results. In particular, the reported results of two experimental studies, which addressed the formation of aquatic ripples in closed conduits, are considered here. One of the studies was of Coleman et al. \cite{Coleman_1}, who experimentally studied bed instabilities in a $6\,m$ long horizontal closed conduit with a rectangular cross section ($300\,mm$ wide by $100\,mm$ high) and employed water as the fluid medium and glass beads as the granular medium. The experiments were performed in a fully turbulent regime, with Reynolds numbers $Re=UH/\nu$ in the range $26000 < Re < 70000$ (where $H$ is the channel height, $\nu$ is the kinematic viscosity, and $U$ is the mean velocity of the fluid). The other study considered here is of Franklin \cite{Franklin_3}, who experimentally studied initial instabilities in a $6\,m$ long horizontal closed conduit with a rectangular cross section ($120\,mm$ wide by $60\,mm$ high). The experiments were performed in a fully turbulent regime, with Reynolds numbers within the range $13000 < Re < 24000$.

\begin{figure}
   \begin{minipage}[c]{.49\linewidth}
    \begin{center}
     \includegraphics[width=\linewidth]{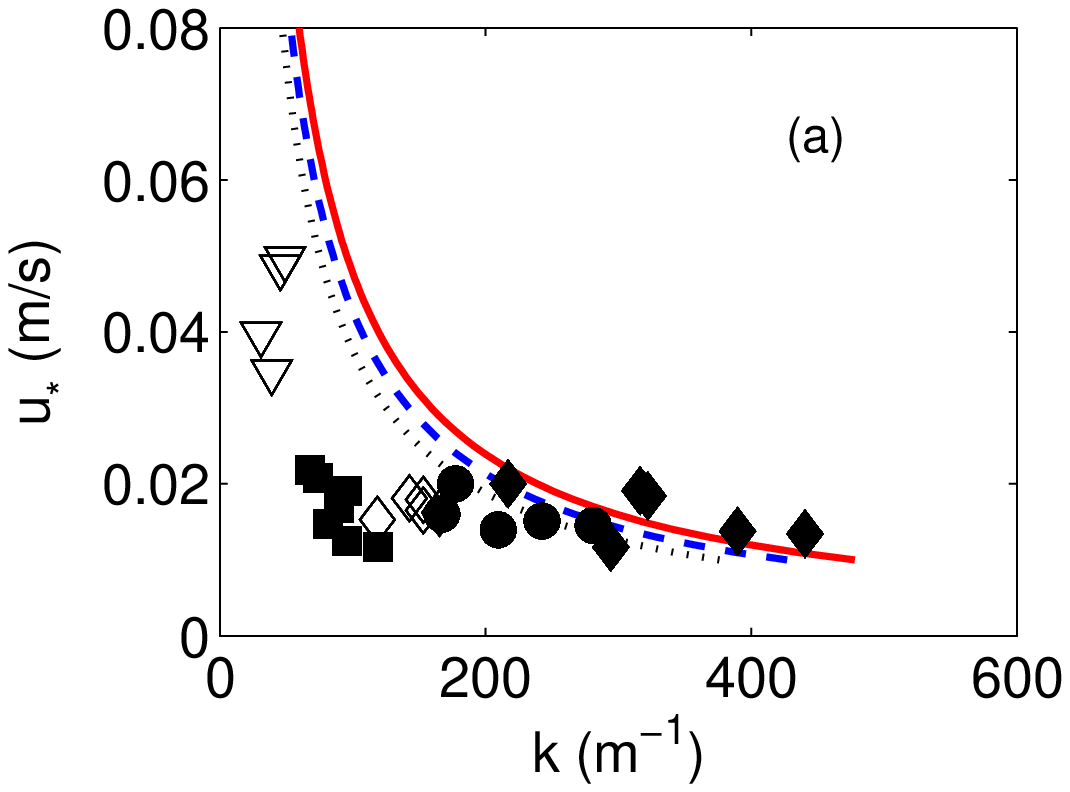}
    \end{center}
   \end{minipage} \hfill
   \begin{minipage}[c]{.49\linewidth}
    \begin{center}
      \includegraphics[width=\linewidth]{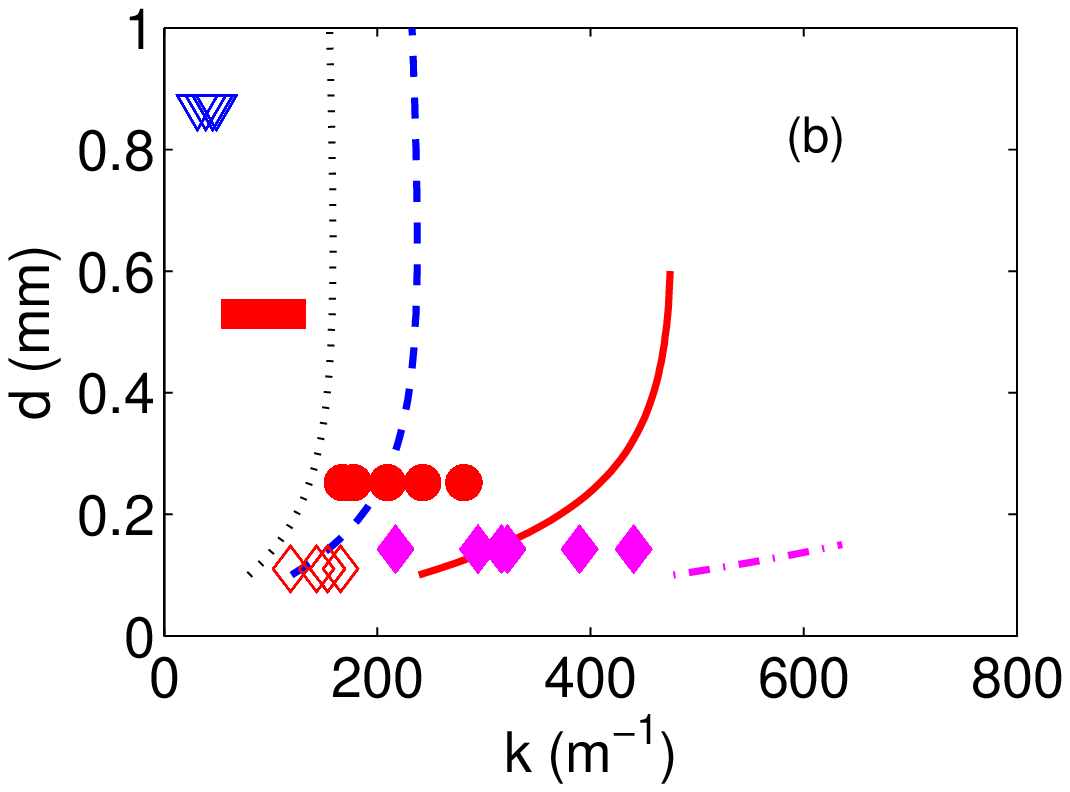}
    \end{center}
   \end{minipage}
\caption{(a) Curves of marginal stability in terms of $u_*$ for different values of $B_g$. The continuous, dashed and dotted curves correspond to $B_g=0$, $B_g=0.01$, and $B_g=0.02$, respectively. (b) Curves of marginal stability in terms of $d$ for different values of $u_*$, where all the parameters affected by $d$ were also varied. The dashed-dotted, continuous, dashed, and dotted curves correspond to $u_*=0.01\,m/s$, $u_*=0.02\,m/s$, $u_*=0.04\,m/s$, and $u_*=0.06\,m/s$, respectively. The symbols correspond to experimental data and are described in the text.}
\label{fig:comp_exp}
\end{figure}

Figure \ref{fig:comp_exp} shows a comparison of the curves of marginal stability obtained in this study with the corresponding experimental data of Coleman et al. \cite{Coleman_1} and Franklin \cite{Franklin_3}. In this figure, the open diamonds and inverted triangles, taken from Coleman et al. \cite{Coleman_1}, correspond to $d=0.11\,mm$ and $d=0.87\,mm$ glass spheres, respectively, under different shear velocities. The filled diamonds, circles, and squares, taken from Franklin \cite{Franklin_3}, correspond to glass spheres with $d=0.14\,mm$, $d=0.25\,mm$, and $d=0.53\,mm$, respectively, under different shear velocities.

Figure \ref{fig:comp_exp}(a) shows the curves of marginal stability in terms of $u_*$ for different values of $B_g$, where the continuous, dashed, and dotted curves correspond to $B_g=0$, $B_g=0.01$, and $B_g=0.02$, respectively. The model results are in agreement with experimental data, since all the measured ripples, with the exception of a few ripples with $d=0.14mm$, lie in the predicted unstable regions. The measured ripples that are not in the unstable region are close to the marginal curves. This discrepancy may be related to experimental uncertainties, but a more probable reason pertains to some model parameters whose values are not well known. One of them is $C_{sat}$, whose value has not yet been measured in the aquatic case, and the other parameter is the threshold shear. In the present analysis, the Shields parameter at the bed-load threshold, i.e., $\theta_{th}=u_{*,th}/\left( (S-1)gd\right)$, where $u_{*,th}$ is the corresponding shear velocity \cite{Bagnold_1}, was fixed at $0.04$. However, there is no real consensus about this value, which may vary between $0.02$ and $0.06$ in the present case \cite{Mantz_1, Yalin_2, Soulsby_Whitehouse, Buffington_1}.

Figure \ref{fig:comp_exp}(b) shows the curves of marginal stability in terms of $d$ for different values of $u_*$, where all the parameters affected by $d$ were varied accordingly. The dashed-dotted, continuous, dashed, and dotted curves correspond to $u_*=0.01\,m/s$, $u_*=0.02\,m/s$, $u_*=0.04\,m/s$, and $u_*=0.06\,m/s$, respectively. Considering that the open diamonds and inverted triangles correspond to $0.015\,m/s<u_{*}<0.019\,m/s$ and $0.034\,m/s<u_{*}<0.050\,m/s$, respectively, and that the filled diamonds, circles, and squares correspond to $0.011\,m/s<u_{*}<0.020\,m/s$, $0.012\,m/s<u_{*}<0.021\,m/s$, and $0.011\,m/s<u_{*}<0.022\,m/s$, respectively, it can be confirmed that all the measured ripples lie in the predicted unstable regions and that the curves of marginal stability obtained in this study agree well with the published experimental data.

As a final remark, I note that previous works proposed different behaviors for the wavelength of aquatic ripples with respect to fluid flow conditions. For instance, Coleman et al. (2003) \cite{Coleman_1} proposed that the wavelength does not vary with the applied fluid flow, whereas Charru et al. (2013) \cite{Charru_5} proposed that the wavelength may decrease with the fluid flow. The present paper presented a linear stability analysis of a sheared granular bed, taking into consideration all the main mechanisms and parameters involved in the turbulent liquid case. Although the most unstable wavelength increases with both the shear stress and the grains diameter, the combination of different mechanisms changes the wavelength selection. In some cases, a granular bed under a given shear stress may develop smaller ripples than a bed submitted to a lower shear stress due to the combination of all the mechanisms and parameters. In other cases, the wavelength may be insensitive to fluid flow conditions. This is not in disagreement with previous works \cite{Fourriere_1, Charru_5, Coleman_1}, especially if we take into consideration the uncertainties involved in the determination of the threshold Shields number (bed armouring included).

\section{Conclusions}
\label{section:conclusions}

This paper addressed the formation of sand ripples under turbulent flows of liquids. It presented a linear stability analysis of a granular bed below a turbulent boundary layer, taking into consideration all the main mechanisms and parameters involved in the turbulent liquid case. In particular, the bed compactness and bed-load threshold shear stress--two parameters that have not been exhaustively considered in previous analyses--were taken into account. The stability analysis showed that the bed compactness does not influence the wavelength, and as its variation is small in practical situations, its influence on the growth rate and celerity may be neglected. In the case of liquids, gravity effects are small and they may also be neglected (as is usually the case in many of the previous analyses). On the other hand, fluid flow and threshold effects have strong influences on the wavelengths, growth rates, and celerities, and therefore, they must be taken into account. Unlike previous stability analyses, in this analysis, expressions were proposed considering all these effects and curves of marginal stability were plotted. The curves of marginal stability were compared with published experimental results and they were found to agree well.

\section{Acknowledgments}

\begin{sloppypar}
The author is grateful to FAPESP (grant no. 2012/19562-6), to CNPq (grant no. 471391/2013-1) and to FAEPEX/UNICAMP (conv. 519.292, project AP0008/2013) for the provided financial support.
\end{sloppypar}






\bibliography{references}
\bibliographystyle{elsart-num}







\end{document}